\documentstyle[12pt,epsfig]{article}
\author{ A. Segu\'{\i}
\footnote{segui@unizar.es}\\
Departamento de F\'{\i}sica Te\'orica. Facultad de Ciencias.\\
Universidad de Zaragoza. 50009-Zaragoza, Spain}
\title{A toy model for the coincidence problem
\footnote{\it I dedicate this research to the memory of 
my thesis advisor Angel Morales}}

\begin{document}
\maketitle
\begin{abstract}
The measured values of the matter energy density and the 
vacuum energy density are obtained using an adiabatic black hole
percolating model at the critical point. The percolation of black holes 
is related with an expanding isotropic universe filled 
with the most entropic fluid saturating the holographic bound. 
\end{abstract}

\vspace{2cm}

KEYWORDS: cosmological constant, black holes, holography, string/M theory.

\vspace{.5cm}
PACS : 98.80.Es,04.70.Bw, 04.20.Gz, 11.25.Yb
\vspace{.5cm}

It has been a surprise that the recent results reported by the WMAP and others 
teams (\cite{wmap}) converge to a universe with an important 
contribution to the energy 
content in the form of vacuum energy. This observation acutes the cosmological 
constant problem; now not only the value of the cosmological constant is 
very small  with respect to the theoretical predictions
but it is actually non-zero and of the order of the 
energy density of ordinary matter 
(dark and visible). Usually one believes that there is some unknown mechanism 
originating on unknown symmetries of the full quantum gravity theory which 
can give a 
cosmological constant equal to zero. But the fact that the cosmological 
constant is very small and also of the same order of the rest of the energy,  
that evolves with the cosmological evolution, is a paradox for which we 
have no mechanism that can explain this phenomenon.

In this research we address this so-called {\it coincidence} 
problem (\cite{weinberg}) by 
envisaging a toy model with the desired properties, and related to the 
vacuum structure of the effective theory. In fact the main problem of the 
string physics is the nature of the vacuum of the 
theory (\cite{vacuum}); in spite that different string theories 
are related by dualities pointing to one single underlying structure, 
the vacuum is not fixed and on the contrary, a great number of 
vacua are in principle possible;  there is not a 
preferred vacuum and due to such a high diversity actually it begins to 
increase the suspition that arguments beyond the theory such as the anthropic 
principle must be advocated to understand the observations(\cite{anthropic}).

In fact the observations must be the clue to gain insight on the nature of the 
vacuum of the string/M theory, and observation, 
mainly cosmological measurements 
offers us with the surprise of a flat, accelerated universe with a 
contribution of vacuum energy, and ordinary energy of the same order. It is 
not easy to relate these observations to the fundamental theory of quantum 
gravity but they must be a consequence of the classical effective theory. 
That is, the string/M theory has to have a vacuum structure that in the 
classical regime must explain the observed cosmological data.

We propose a mechanical model to account for the cosmological observations 
of the composition of the universe; 
we consider a fluid of black holes evolving 
adiabatically as the substratum behind the actual universe. A fluid of black 
holes has been considered before by T. Banks and W. Fischler as the initial 
state for the Universe (\cite{BF}) 
because such a fluid saturates the holographic bound(\cite{holo}) 
underlying a holographic cosmology (\cite{cosmo}). We insist, 
with variants, on this mechanical model to account for the actual observations.

We consider a system of black holes in a process of coalescence. We
neglect the gravitational interaction between them so that the process is
purely random; as the density of black holes increase the system approach the 
critical point at which it percolates. We consider the vicinity of 
this critical point when the infinite cluster develops; then we identify
a cosmological model with the infinite cluster of a network of 
percolating black holes (\cite{perco}). 

This model at the critical point shares many properties with the dense black 
hole fluid of (\cite{BF}); in fact the infinite cluster or infinite 
droplet will be a black hole covering all the disposable space. We 
envisage a dual description for the system. For an observer
\footnote{We use the term observer without any anthropomorphic intention; 
the geometry is independent of the presence of any observer.}
surfing the expanding event horizon the geometry is that of a
Friedman Lema\^{\i}tre Robertson Walker 
(FLRW) universe filled with a fluid 
with equation of state $p=\omega \rho$ with $\omega=1$ (\cite{BF}). 
For an observer that
traverses the event horizon the geometry is the interior of a growing 
black hole; in its expansion the density of matter is diluted and 
asymptotically tends to zero as the radius of the black hole goes to 
infinity; the geometry is described by a flat expanding FLRW model.

\begin{figure}[!hbt]
\begin{center}
\includegraphics[width=10cm]{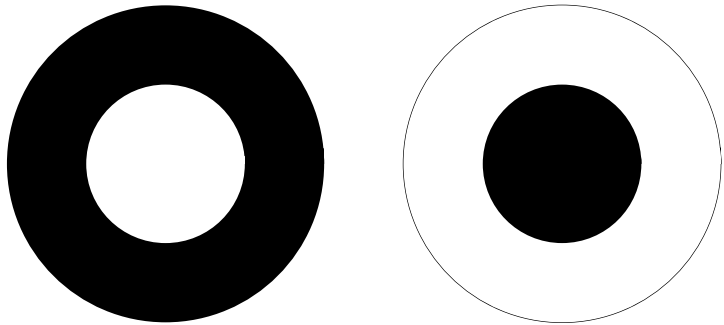}
\end{center}
\caption{{\small 
The two dual possibilities are sketched. In the right the horizon of a 
black hole is surrounded by matter. In the left ordinary matter is covered
by an horizon of the percolating black hole system.
}}
\label{bh}
\end{figure}

At the percolating critical point there is a duality between the regions 
covered by black holes, that are inside an event horizon (black regions), 
and regions outside the event horizon (white regions)(Figure 
\ref{bh}). In (\cite{BF}) 
the white regions are used to contact with the observational cosmology. 
We need to use the Israel Junction conditions (\cite{Israel}) to match 
the geometry at both sides of the event horizon for this interstitial regions.
What it is evident is that this white regions are covered by an event horizon 
and are candidates to develop accelerated cosmologies.

Our toy model of a percolating network of black holes at the critical point
can describe two different scenarios. In the black regions the expansion 
is decelerated, the matter is diluted as the infinite cluster grows, and 
asymptotically the density is zero ( the relation between the density and 
size for a black hole is $\rho \sim R^{-2}$). This realization of the 
model is a candidate for the initial expansion of the universe; we have 
a FLRW expanding model isotropic and flat without an initial singularity
\footnote{We have not the big bang singularity but because we are in 
the interior of black holes we confront the black holes singularities; 
however if the time of merging is smaller than the time of collapse the 
singularities are not formed and we have a singularity free model.}.
The isotropy and flatness are consequences of the random distribution 
of black holes and the fact that we are at the critical percolating point;
this two properties appear in the model of Banks and Fischler and 
can be related with the saturation of the entropy 
bounds proposed in (\cite{BF}) to construct a holographic cosmology; 
moreover, the percolating critical point is scale invariant and the 
distribution of clusters is described by a power law; we have not been 
able to determine the critical exponents of our model but if the 
relation with the cosmic fluid $p=\rho$ can be made we can use the 
results of (\cite{BF}) to assert that the distribution is a  
Gaussian random scale invariant one.

In the white regions 
we have ordinary radiated matter surrounded by an event horizon; this geometry
has the causal structure of accelerated cosmologies; if a cosmological 
constant is the responsible of the dark energy, the event horizon is at 
a fixed proper distance $ R \sim \Lambda^{-2}$, whereas if the dark energy is 
some sort of quintessence with $-1/3 > \omega > -1$, the  event horizon 
grows with the cosmic time at a rate $R \sim t$, $t$ being the cosmic time;
so in this regions covered by an event horizon the expansion 
can be accelerated. Is
in these last regions that we make contact with the observed universe.
With this end we analyze the merging of black holes.

When two black holes meet due to the contact of its event horizons they form a 
new black hole
\footnote{ When we say black hole we refer to a closed event horizon; 
it is clear that the geometry is strongly distorted with respect to the 
Schwarzschild one}.
We consider two magnitudes in this process, the energy and the entropy. For 
the energy conservation law we have
 \footnote{
We use units in which $G=c=\hbar=k_B=1$. To restore the correct dimensions
on a formula multiply by the appropriate powers of the previous constants.}
\begin{equation}\label{1}
2M=M_{f}+E_r,
\end{equation}
where $M$ is the mass of each of the initial black holes, that we suppose the 
same, $M_{f}$ is the mass of the final black hole and $E_{r}$ is the
energy emitted in the process of merging; using the relation $R=2M$ between 
radius and mass for Schwarzschild black holes we have
\begin{equation}\label{2}
R={R_{f} \over 2 } + E_{r}.
\end{equation}

On the other hand, the second law of thermodynamics establishes that the 
entropy of the final state is equal or higher than the initial one; for 
black holes the entropy is $1 \over 4$ the area of the event horizon in
Planck units, so we have
\begin{equation}\label{3}
2R^{2} \leq R_{f}^{2} + {S_{r} \over \pi},
\end{equation} 
where $S_{r}$ is the entropy of the radiated energy $E_{r}$. A good 
approximation is to consider $S_{r} \ll R_{f}^{2} \sim R^{2}$ so that for
an adiabatic evolution we have
\begin{equation}\label{4}
2R^{2} =R_{f}^{2}.
\end{equation}
In this adiabatic coalescence of two equal mass black holes we finish with
to forms of energy, one constituting a final black hole and other in form of
radiated energy. The amount of energy in the final black hole with respect of
the total energy $2M$ can be obtained from (\ref{4}),
\begin{equation}\label{5}
\Omega_{bh}={M_{f} \over 2M}= {R_{f}/2 \over R}={ 1 \over \sqrt{2} }=.707;
\end{equation}
consequently, the percentage of energy radiated away during this process is 
$\Omega_{r}= E_{r}/ 2M =  1 -1/ \sqrt{2} =.293$ (\cite{Hawking}). 
This two forms of energy, one associated
to the event horizon of a final black hole and the other to ordinary energy,
coincide with the two contributions to the energy of the universe reported by
observations (\cite{wmap}) in terms of an accelerating amount of energy 
$\Omega_{\Lambda} \sim 0.7$, and a decelerating one 
$\Omega_{m} \sim 0.3$. 

We can relax the equality of the initial black hole masses; if we consider the 
union of two black holes of mass $M$ and $\alpha M$ with the same premises 
as before we can show that the amount of radiated energy is given by
\begin{equation}\label{6}
\Omega_{r}=1-{ \sqrt{1+\alpha^{2}} \over 1+\alpha},
\end{equation}
that takes the maximum value for 
$\alpha=1$. Also it is clear that without radiated energy ($E_{r}=0$), 
the conservation of energy (\ref{1}) and the conservation of entropy (\ref{4})
are not compatible. 

If we consider the entropy of the radiated energy $S_{r}$ in the process of 
adiabatic coalescence between two equal mass 
black holes, our model allows us to 
relate the properties of the matter radiated (which according to 
the observation is made mainly by dark matter) with the 
value of $\Omega_{\Lambda}$. Manipulating (\ref{1}) and (\ref{3}), 
this last with an equal sign, we obtain, 
\begin{equation}\label{7}
{E_{r}^{2} \over S_{r}} = {1 \over 2 \pi } {(1-\Omega_{\Lambda})^2 \over 
1-2 \Omega_{\Lambda}^{2}}.
\end{equation}

Let us now elucidate the loopholes that the previous model has. In the 
case in which the formation of a single black hole 
is given by the coalescence of an
arbitrary number $n$ of initial ones we obtain $\Omega_{bh}=1/\sqrt{n}$; so
the model accords with observation only if the dominant channel is the
approximately equal mass pairing. A sort of bootstrap model can be
envisaged in which the merging process is due to the contact of 
closed event horizons favoring the single pairing as the elementary mode.
If the adiabatic merging of pairs of equal mass black 
holes is the elementary process of the percolation of black holes  
the percolation process will share the same partitioning for the 
total energy as the elementary one.

Our model predicts a constant relation between $\Omega_{\Lambda}$ and
$\Omega_{m}$ during the process of percolation. The pairing process 
extends to all spatial scales and consequently has in principle 
an infinite duration in time. In order to respect primordial nucleosynthesis
we must put an initial cosmic time for the beginning of the percolation process
\emph {after} primordial nucleosynthesis be developed in the standard 
way. There is an scale in this approach given by the size of the 
initial black hole that we can relate with the 
observational indications of a possible cutoff in the spectrum
of multipoles of the cosmic microwave background radiation at 
low momentum (\cite{low}).

Despite the previous problems the black hole fluid is a very interesting
sort of vacuum of the underling theory of quantum gravity. It 
saturates the entropy bounds, so it is the more 
entropic (and consequently anthropic) fluid we can construct. Also it is a
good candidate for the initial state of the universe, 
and we have shown that using its properties we can
explain the observed values for the two sort of energies that we observe, with
simple assumptions. It is a vacuum that seems to have in its nature 
different realization for accelerated and decelerated cosmologies. We think
it merits a deeper study.

\section*{Acknowledgements} \nonumber
I thank R.~Emparan for helpful comments. I thank also L.~J.~Boya for reading
the manuscript. This work was supported by MCYT (Spain), grant FPA2003-02948.

\end{document}